\documentclass[12pt]{iopart}

\usepackage{graphicx}

\RequirePackage{color}

\begin{document}

\title[Modeling compact stars]{Modeling compact stars without numerical integration}

\author{Hilario Rodrigues\footnote{E-mail: harg@cefet-rj.br}}

\address{Departamento de F{\'i}sica \\
 Departamento de F\'isica, Centro Federal de Educa\c{c}\~ao Tecnol\'ogica/RJ \\
Av. Maracan\~a 229, 20271-110, Rio de Janeiro, RJ, Brazil}

\begin{abstract}
Taking a novel approach, this paper discuss the structure of compact stars, an important topic in theoretical astrophysics. Adopting the Newtonian gravitation, we solve the hydrostatic equilibrium equation by imposing a simple parametrization for the mass density inside the star. The solutions of the equilibrium equation are carried out without numerical integration, with the aim of determining few global properties of white dwarfs and neutron stars. The global properties of compact stars are thus provided by simple algebraic relationships. The model is intended as an introductory approach to the study of compact stars at the undergraduate or graduate levels.
\end{abstract}

\maketitle

\section{Introduction}

White dwarfs and neutron stars - compact stars for short - are a very important topic of the branch of relativistic astrophysics. One interesting aspect concerns to the basic mechanisms of formation of these objects, which still are not fully understood,  mainly the neutron stars. Indeed, white dwarfs and neutron stars are the endpoint of the evolution of moderately massive stars.  Due to high densities reached in their interior, relativistic effects are important ingredients to be included in the study of such class of stars.

 The study of compact star structure is usually treated either by complete resolution of structure equations, or presenting simplified argument based upon an approximate integration of the hydrostatic equilibrium equation, and the equation of state of the stellar matter, which leads to the derivation of the $M$ - $R$ relation of the compact star \cite{Salaris,BH,Clayton}.

The analysis of the structure of compact stars consists basically of the solution of the Lane-Emden equation (for white dwarfs) and the Tolman-Oppenheimer-Volkoff equation (for neutron stars), which are ultimately the hydrostatic equilibrium equations derived from the Newtonian gravitation and  from the General Relativity, respectively. In the latter case, the general-relativistic effects are taken into account considering the Schwarzschild metric valid for a time-independent and spherically symmetric mass distributions \cite{BH,Glend,Fridolin}.  

Because the difficulty of the involved concepts and numerical methods, such a topic is generally treated in advanced physics courses. Thus, it may be interesting propose new approaches on this topic to students of science and engineering at undergraduate level, who want a first contact with the subject.

In this work, we present a simple model for calculating the structure of compact stars where the density inside the star is parameterized as a decreasing function of the distance from the center of the star. We assume valid the Newtonian gravitation in describing the equilibrium properties of compact stars. The model provides the mass and the radius of the star, for a given fixed value of the central density of the star. The obtained results are partially comparable to those provided by the more realistic but more complex models.

The work is addressed mainly to undergraduate students and teachers. The study of the topic requires the domain of calculus and several  concepts from various fields of physics at intermediate level. 

\section{Description of the model}

We start assuming that the distribution of matter inside the star has spherical symmetry and follows  the power law
\begin{equation}
\rho (r) = \rho_c \left(1 - \frac{r^2}{R^2} \right)^n, \label{1}	
\end{equation}
where $\rho_c$ is the central density, $ r$ is the distance from the center of the star, $R$ is the radius of the star, and $n$ is a positive integer index.  According to equation (\ref{1}), the density of matter inside the star is a monotonically decreasing function of the distance from the center, which vanishes on the surface of the star, as required \cite{Salaris,BH}. This choice is not unique, but we adopt this form in this work for the sake of simplicity.  
   
 In the framework of the Newtonian Gravitation Law, the equilibrium of the star is governed by the equation
\begin{equation}
\frac{dp}{dr} = - \frac{G \rho (r)  m(r)}{r^2}, \label{eql}	
\end{equation}
where  $p$ is the pressure, $G$ is the gravitational constant, and
\begin{equation}
 m(r) = \int_0^r 4 \pi \rho {r^{\prime}}^2 dr^{\prime} \label{2}	
\end{equation}
is the mass enclosed within the sphere of radius $r$. Essentially, the equilibrium equation written above says that the inward force due to gravity  inside the star is counterbalanced by the outward hydrostatic force due to the medium pressure. Using the density law in equation (\ref{1}), we obtain the following expression for the enclosed mass:
\begin{equation}
m(x) = \frac {4 \pi}{3} R^3  x^3 \rho_c \,  _2 F _1 \left (\frac{3}{2},-n;\frac{5}{2};x^2 \right ), \label{massr} 
\end{equation}
where the parameter $x$ is defined by $x = r/R$ ($0 \le x \le 1$), and $_2 F _1 (\cdot) $ denotes the hypergeometric function, which for   $\left| z \right| <1$ can be defined  by \cite{Abramo1,Abramo2}
\begin{equation}
	_p F _q (a_1,\cdots,a_p;b_1,\cdots,b_q;z) = \sum_{n=0}^{\infty} \frac{(a_1)_n\cdots(a_p)_n}{(b_1)_n\cdots(b_q)_n} \frac{z^n}{n!}.
\end{equation}
In the last equation $(q)_n$ stands for the Pochhammer symbol, which for $n>0$ is defined by \cite{Abramo1,Abramo2} 
\begin{equation}
	(q)_n \equiv  \frac{\Gamma \left( q + n \right)}{\Gamma(q)} = q (q+1)(q+2) \cdots (q+n-1),
\end{equation}
where $\Gamma (\cdot)$ represents the gamma function \cite{Abramo2}.

\begin{table}[htbp]
\begin{center}
\caption{\label{table1} Enclosed mass as a function of $x=r/R$. }
\vspace*{.3cm}
\begin{tabular}{c|l}
\hline
 Index $n$ &  $m (x)$ in units of $4 \pi R^3 \rho_c$ \\
\hline
 1 & $-\frac{1}{5} x^5 + \frac{1}{3} x^3 $ \\ 
 2 & $ \frac{1}{7} x^7-\frac{2}{5} x^5 + \frac{1}{3} x^3 $\\
 3 & $ - \frac{1}{9} x^9 + \frac{3}{7} x^7-\frac{3}{5} x^5 + \frac{1}{3} x^3 $\\ 
 4 & $ \frac{1}{11} x^{11} - \frac{4}{9} x^9 + \frac{6}{7} x^7-\frac{4}{5} x^5 + \frac{1}{3} x^3 $ \\
 5 & $ - \frac{1}{13} x^{13} + \frac{5}{11} x^{11} - \frac{10}{9} x^9 + \frac{10}{7} x^7- x^5 + \frac{1}{3} x^3 $ \\
\hline
\end{tabular}
\end{center}
\end{table}

For $n = 2$, for example, one obtains for the enclosed mass 
\begin{equation}
	m(x) = {4 \pi}  R^3 \rho_c \left( \frac{x^7}{7} - \frac{2 x^5}{5} + \frac{x^3}{3} \right). 
\end{equation}

For the sake of illustration, in Table \ref{table1} we show the expressions of the enclosed mass as functions of $x$ for some values of the index $n$.  There is one interesting thing to be pointed out in Table \ref{table1}: if we put $x =1$, i.e. if we consider the outer radius of the star, the enclosed mass becomes the total mass and its value, expressed in units of $4\pi R^3 \rho_c$, decreases as the index $n$ grows up. As this value is also equal to the ratio of the average to the central density, this entails that the stellar configuration becomes more centrally-condensed as $n$ increases. The same behaviour occurs for polytropic configurations, although the equation of state of degenerate gas -- that has a polytropic form -- has not yet been introduced. This behavior in fact comes from Eq. (\ref{1}).

Integrating equation (\ref {2}) from $r=0$ ($x=0$) to $r=R$ ($x = 1$), provides the closed expression of the total mass $M$ of the star, which reads
\begin{equation}
	M = \pi^{3/2}  R^3 \rho_c \frac{\Gamma (n+1)}{\Gamma \left( n + \frac{5}{2} \right)}. \label{mass21} 
\end{equation} 

Inserting (\ref{1}) and (\ref{2}) into equation (\ref{eql}) and carrying out the integration, we get the pressure as a function of $x$:
\begin{equation}
	p(x) = p_c - \frac {4 \pi}{3} G  R^2 \rho_c^2 \int x (1-x^2)^n \, _2 F _1 \left (\frac{3}{2},-n;\frac{5}{2};x^2 \right ) dx, \label{px}
\end{equation}
where  $p_c$ is the pressure at the center of the star. In Table \ref{table2} we show the obtained expressions of the pressure as a  function of $x$ for some values of the index $n$.

\begin{table}[htbp]
\begin{center}
\caption{\label{table2} Pressure as a function of $x=r/R$. }
\vspace*{.3cm}
\begin{tabular}{c|l}
\hline
 Index $n$ &  $p_c - p(x)$ in units of $4 \pi G R^2 \rho_c^2$ \\
\hline
 1 & $ \frac{1}{30}x^{6}-\frac{2}{15}x^{4}+\frac{1}{6}x^{2} $ \\ 
 2 & $ \frac{1}{70}x^{10}-\frac{3}{35}x^{8}+\frac{67}{315}x^{6}-\frac{4}{15}x^{4}+\frac{1}{6} x^{2} $\\
 3 & $ \frac{1}{126} x^{14}-\frac{4}{63}x^{12}+\frac{233}{1050}x^{10}-\frac{139}{315}x^{8}+\frac{
113}{210}x^{6}-\frac{2}{5}x^{4}+\frac{1}{6}x^{2} $\\ 
\hline
\end{tabular}
\end{center}
\end{table}

Finalizing, if one imposes the condition that the pressure is zero at the surface of the star, i. e. $p(x=1) = 0$, one obtains 
\begin{equation}
	p_c = \frac {2 \pi G }{3(n+1)} R^2 \rho_c^2  \, _2 F _1 \left (1,\frac{3}{2},-n;\frac{5}{2},n+2;1 \right ). \label{press} 
\end{equation}

 Generally, the equations (\ref{1}), (\ref{massr}) and (\ref{px}) stand for the stellar structure equations. In addition, the following boundary conditions on the surface of the star -- that is in the limit $x=1$ -- are needed: $m(x=1)=M$, $p(x=1)=0$ and $\rho(x=1)=0$. As a whole, the free quantities introduced in the model are: the central pressure $p_c$, the central density $\rho_c$, the stellar  radius $R$, and the stellar mass $M$. An equivalent way to see this is to resort to equation (\ref{mass21}), from which we obtain a $M$-$R$-$\rho_c$ relation, and to equation (\ref{press}), that provides a $p_c$-$R$-$\rho_c$ relation.  Overall, we have four free quantities and two relations, i.e., only two degrees of freedom left.  

  This entails that we can choose as free independent quantities, in principle, both $M$ and $R$ and hence we do not have any $M$-$R$ relation. In fact, a $M$-$R$ relation for white dwarfs follows only when another relation is added. In the next section we introduce the equation of state for degenerate gas relating the pressure and the density inside the white dwarf. This hence eliminates a degree of freedom and yields a $M$-$R$ relation. As we will see, the constant $K$ in equation (\ref{polytropic}) below is not a generic constant but it is well determined by physical theory: this is the reason why an actual $M$-$R$ relation for white dwarfs occurs. If it were an undetermined constant, like it happens when introducing a polytropic description of e.g. a radiative star in ideal gas condition, then the degrees of freedom would remain two and no $M$-$R$ relation would follow.

\section{White dwarfs}

Let us first consider the application of the model described in the last section to the study of white dwarfs. In few words, white dwarfs are objects composed basically of ionized atomic nuclei wrapped in a highly degenerate electron gas, with mass typically between $0.5$ $M_\odot$ and $1.45$ $M_\odot$, where $M_\odot =2.0 \times 10^{33}$ g is the solar mass. The electron pressure is high enough to maintain the equilibrium of the star, counterbalancing the gravitational attraction \cite{Salaris,BH,Clayton}. 

  Equation (\ref{press}) is the key to determine the star structure. Only we need is the value of the pressure at the center of the star for a given value of the central density $\rho_c$. To fixing the central pressure in equation (\ref{press}), it is customary to suppose that the pressure and the density are related by the polytropic equation of state
\begin{equation}
	p = K \rho^\gamma , \label{polytropic}
\end{equation}
where $K$ is a constant and $\gamma$ is the characteristic adiabatic index of the star.  Of course, the same equation holds in particular at the centre of the star. Thus, inserting this equation on the left hand side of equation (\ref{press}), we obtain the radius as a function of the central density,
\begin{equation}
R=\left[ \frac{3\left( n+1\right) K}{2\pi G}\right] ^{1/2}\rho _{c}^{\frac{%
\gamma -2}{2}} \,  _{2} F_ {1}^{-1/2} \left( 1,\frac{3}{2},-n;\frac{5}{2}%
,n+2;1\right)  . \label{radius1} 
\end{equation} 

Replacing $R$ in equation (\ref{mass21}) we have for the mass of the star
\begin{eqnarray}
M =&& \left[ \frac{3\left( n+1\right) K}{2G}\right] ^{3/2}\rho _{c}^{\frac{%
3\gamma -4}{2}}\frac{\Gamma \left( n+1\right) }{\Gamma \left( n+\frac{5}{2} \right) } \nonumber \\
&& \times  \, _{2} F _{1} ^{-3/2}\left( 1,\frac{3}{2},-n;\frac{5}{2},n+2;1\right)  . \label{mass31}
\end{eqnarray} 

 As $M$ and $R$ are observable quantities, while $\rho_c$ is not, hence an explicit $M$-$R$ relation is preferable. Eliminating $\rho_c$ in the last equation by using (\ref{radius1}), we arrive at
\begin{eqnarray}
M=&& \pi ^{\frac{3\gamma -4}{2\gamma -4}}\left[ \frac{3\left( n+1\right) K}{2G}%
\right] ^{-\frac{1}{\gamma -2}}R^{\frac{3\gamma -4}{\gamma -2}}\frac{\Gamma
\left( n+1\right) }{\Gamma \left( n+2.5\right) } \nonumber \\ 
&& \times \,  _{2} F_ {1} ^{\frac{1}{\gamma -2}} \left( 1,%
\frac{3}{2},-n;\frac{5}{2},n+2;1\right)  . \label{mass41}
\end{eqnarray}

For low densities ($\rho \ll 10^6 \,{\rm g \cdot cm}^{-3}$ ), the electron gas is well described as a non-relativistic Fermi gas, for which $\gamma = 5/3$. In this case, the above expressions for the radius and the mass of the star
are thus given by 
\begin{equation}
R=\left[ \frac{3K\left( n+1\right) }{2\pi G}\right] ^{\frac{1}{2}}\rho
_{c}^{^{-\frac{1}{6}}}\ _{2}F_{1}^{-\frac{1}{2}}\left( 1,\frac{3}{2},-n;%
\frac{5}{2},n+2;1\right) , \label{rnrwd}
\end{equation}
and
\begin{equation}
M=\pi ^{-\frac{3}{2}}\left[ \frac{3K\left( n+1\right) }{2G}\right]
^{3} \frac{\Gamma \left( n+1\right) }{\Gamma \left( n+\frac{5}{2}%
\right) } R^{-3} \ _{2}F_{1}^{-3}\left( 1,\frac{3}{2},-n;\frac{5}{2},n+2;1\right) . \label{massrm3} 
\end{equation} 

The preceding results entails that in the non-relativistic limit the radius (mass) of the white dwarf behaves as a decreasing (increasing) function of the central density. We can see also that the model reproduces the well-known $M \propto R^{-3}$ dependence, which is usually the result of a full white dwarf treatment in the limit of low central density.
 
For the sake of example, if we put $n=2$ we obtain 
\begin{equation}
R	= \alpha_2^{1/2} \rho_c^{-1/6} , \label{radius12}
\end{equation}     
and the mass
\begin{equation}
M = \frac{32 \pi}{105} \alpha_2^{3/2} \rho_c^{1/2}, \label{mass2}
\end{equation}  
as functions of the central density, where
\begin{equation}
 \alpha_2 = \frac{315 K}{52 \pi G} .  \label{alpha1}
\end{equation}
Eliminating $\rho_c$ between the equations (\ref{radius12}) and (\ref{mass2}), one arrives at
\begin{equation}
M=\frac{32 \pi }{105} \alpha_2^{3} R^{-3}.
\end{equation}
 
 Of course, the dependences $R \propto \rho_c^{-1/6}$,  $M \propto \rho_c^{1/2}$ and $M \propto \rho_c^{-1/3}$ are maintained for different values of the index $n$.

As the density inside the white dwarf increases, the electron energy is no longer negligible relative to the rest mass. The typical velocities of the electrons approach the speed of light, and special relativity effects must be taken into account. Inside a white dwarf, this situation prevails for densities $\rho \gg 10^6 \,{\rm g \cdot cm}^{-3}$, and so in the ultra-relativistic limit the adiabatic index approaches the value $\gamma =  4/3$ \cite{BH}.  In this limit, the model star radius, given by (\ref{radius1}), reads
\begin{equation}
R=\left[ \frac{3K\left( n+1\right) }{2\pi G}\right] ^{\frac{1}{2}}\rho
_{c}^{^{-\frac{1}{3}}}\ _{2}F_{1}^{-\frac{1}{2}}\left( 1,\frac{3}{2},-n;%
\frac{5}{2},n+2;1\right). \label{urr}
\end{equation}
 On the other hand, from equation (\ref{mass41}) we see that the star mass is given by
\begin{equation}
M=\left[ \frac{3K\left( n+1\right) }{2G}\right] ^{\frac{3}{2}}\frac{\Gamma
\left( n+1\right) }{\Gamma \left( n+\frac{5}{2}\right) }\ _{2}F_{1}^{-\frac{3%
}{2}}\left( 1,\frac{3}{2},-n;\frac{5}{2},n+2;1\right) . \label{urr1234}
\end{equation}

 Again, putting $n=2$ for the sake of example, we get 
\begin{equation}
R	= \alpha_2^{1/2} \rho_c^{-1/3} ,\label{radius21}
\end{equation}     
and
\begin{equation}
M = \frac{32 \pi}{105} \alpha_2^{3/2} .\label{mass22}
\end{equation}  

Observe that the model radius is still a decreasing function of the mass density. In contrast, the mass does not depend on the central density $\rho_c$ or the star radius, but solely on the index $n$ and chemical composition. In the two preceding equations, the parameter $\alpha_2$ has the same definition given in equation (\ref{alpha1}). 


 {We can now try to assess the chemical composition of a white dwarf in order to calculate its actual mass for different $n$-values.    Basically, the chemical composition of a white dwarf depends on the mass of the progenitor star and the nuclear stages it went through.  White dwarfs can be made of helium, of a carbon-oxygen mixture, or of a oxygen-neon-magnesium mixture \cite{Salaris,BH,Clayton}. Of course, the information about the chemical composition is carried by the  mean molecular weight per electron $\mu_e$, defined by equation  (\ref{mue}) below. Indirectly, it affects the proportionality constant $K$ of the polytropic equation of state \cite{BH}. 

For the sake of simplicity, in this work we choose only one single composition and carry on the calculations for that case only.  Thus,  considering a white dwarf composed entirely of carbon and a non-relativistic electron gas ($\gamma=5/3$), we have the constant $K=1.0033 \times 10^{13} \mu_e^{-5/3}$, which gives $K = 3.1611 \times 10^{12} \, {\rm g^{-2/3} \cdot cm^4 \cdot s^{-2}}$  \cite{BH}. On the other hand, if the electron gas within the carbon white dwarf is ultra-relativistic ($\gamma=4/3$), then we should use the value $K=1.2435 \times 10^{15} \mu_e^{-4/3}$, which results $K = 4.9348 \times 10^{14} \, {\rm g^{-1/3} \cdot cm^3 \cdot s^{-2}}$ \cite{BH}. 

 Henceforth in this section, we will review some results of the full white dwarf calculations in order to see if our results are comparable to these ones. If the polytropic equation of state $p(\rho) = K \rho^\gamma$ is supposed to hold, the full treatment leads to a second order differential equation for the density, which is known as the Lane-Emden equation \cite{Salaris,BH}. This equation admits only numerical solutions for a given range of values of the adiabatic index $\gamma$.

For low densities ($\rho \ll 10^6 \,{\rm g \cdot cm}^{-3}$) and for $\gamma = 5/3$, the numerical solution of the Lane-Emden equation allows the following form of the star radius \cite{BH,Clayton}:
\begin{equation}
R=\frac{1.9988\times 10^{4}}{\mu _{e}^{5/6}}\left( \frac{\rho _{c}}{10^{6} \,
{\rm g \cdot cm}^{-3}}\right) ^{-1/6}\, {\rm km},  \label{1.48}
\end{equation}%
while the star mass as a function of the radius reads 
\begin{equation}
M=\frac{22.435}{\mu _{e}^{5}}\left( \frac{R}{10^{4}\, {\rm km}}\right)
^{-3} \, M_{\odot }.  \label{1.49.1}
\end{equation}

In the preceding equations, $\mu_e$ stands for the mean molecular weight per electron, which is given by \cite{BH,Clayton}
\begin{equation}
\mu_{e}=\left(\sum_{i}\frac{Z_{i}X_{i}}{A_{i}}\right)^{-1}, \label{mue}
\label{1.47b}
\end{equation}
where $A_i$, $Z_i$ and $X_i$ are the mass number, the atomic number and the fraction of the $i-$th present atomic nuclei,  respectively.

We see from equations (\ref{1.48}) and (\ref{1.49.1}) that in the non-relativistic polytropic approximation the star radius is a decreasing function of the central density, while the star mass is a decreasing function of the cubed star radius.  In general, this behaviour is in  agreement with the results given by equations (\ref{rnrwd}) and (\ref{massrm3}).  

For high densities (($\rho \gg 10^6 \,{\rm g \cdot cm}^{-3}$ ) and $\gamma = 4/3$ (relativistic electron gas) the radius and the mass of the star are given by \cite{BH}
\begin{equation}
R=\frac{5.313\times 10^{4}}{\mu _{e}^{2/3}}\left( \frac{\rho _{c}}{10^{6} \,
{\rm g\cdot cm}^{-3}}\right) ^{-1/3}\, {\rm km},  \label{1.51}
\end{equation}%
and
\begin{equation}
M=1.457\left( \frac{2}{\mu _{e}}\right) ^{2}M_{\odot }.  \label{1.52}
\end{equation}

 Here, we have an interesting result: the radius of the star is still a decreasing function of the central density, but the star mass does not depend on the radius: it is constant. This is in agreement with the results given by equations (\ref{urr}) and (\ref{urr1234}). The constant mass found through the full treatment of the ultra-relativistic white dwarfs is the well-known Chandrasekhar limit \cite{BH,Chandra}.

\begin{figure}[htpb]
\centering 
\includegraphics[width=.25\textheight]{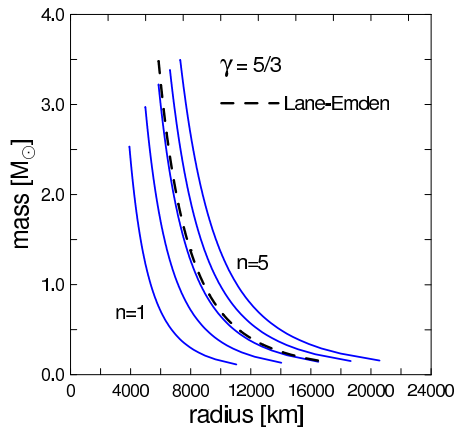}
\vspace*{-.1cm} 
\caption{(Color online) Mass-radius diagrams for the low density model white dwarfs. \label{fig:100}}
\end{figure}

The existence of such a limit means that an extremely relativistic electron gas cannot support a star with mass larger than the Chandrasekhar mass. So, the Chandrasekhar mass is the maximum possible mass for a white dwarf, and when the mass exceeds this limit,  the star collapses. 

Naturally, a more complete treatment of the structure of white dwarfs, even within the proposed model, should interpolate the equation of state $K_1 \rho_c^{5/3}$ for low densities, and $K_2 \rho_c^{4/3}$ for high densities ($\rho \gg 10^6 \,{\rm g \cdot cm}^{-3}$).

Figure \ref{fig:100} shows the mass-radius diagrams obtained with the present model for low density carbon white dwarfs described by a $\gamma = 5/3 $ polytropic equation of state given by equation (\ref{polytropic}) \cite{BH}, and for values of the index $n$ varying from $n=1$ to $n=5$. The $M$-$R$ relation is described by equation (\ref{massrm3}). The dashed line in the figure represent the solutions obtained from the Lane-Emden equation. As we can see, the result obtained with $n=3$ is in good agreement with that obtained with the Lane-Emden equation.

The model radius as a function of the central density is also shown in Figure \ref{fig:200}, for the same set of the parameter $n$, and for central densities in the range  $10^5$ to $ 5 \times 10^7$ ${\rm g\cdot cm}^{-3}$. We see again the same good agreement of the previous figure for $n = 3 $.

\begin{figure}[htpb]
\centering 
\includegraphics[width=.25\textheight]{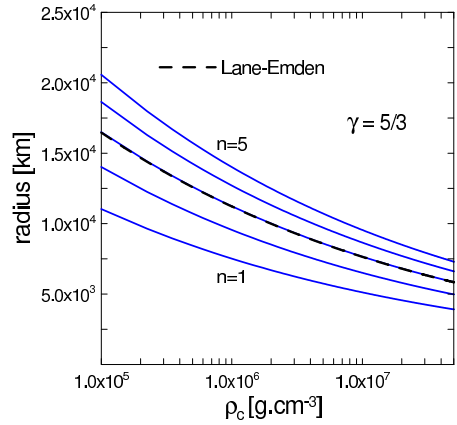}
\vspace*{-.1cm} 
\caption{(Color online) Radius as a function of the central density for low density model white dwarfs. \label{fig:200}}
\end{figure}

\begin{figure}[htpb]
\centering 
\includegraphics[width=.25\textheight]{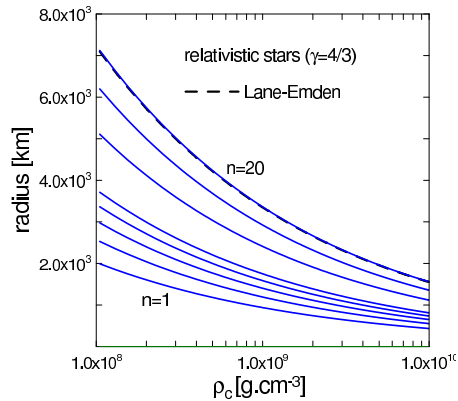}
\vspace*{-.1cm} 
\caption{(Color online) Radius as a function of the central density for high density model white dwarf. \label{fig:300}}
\end{figure}

In Figure \ref{fig:300} we depict the radius of high density white dwarfs as a function of the central density varying in the range  $10^6$ to $ 5 \times 10^8$ ${\rm g\cdot cm}^{-3}$, for few values of the parameter $n$. The star radius is calculated from Eq. (\ref{urr}), with the electrons described as a relativistic $\gamma = 4/3$ polytropic. Note that the dashed line depicted in the figure represent the solutions obtained from the Lane-Emden equations for relativistic white dwarf. In this case the agreement between the results occurs for $n = 20$. Observe that in the high density case only the $ R $-$\rho_c$ relation can be depicted, since no $M$-$R$ relation exists any more.

 In Table \ref{table3} we show the obtained star masses for few values of the parameter $n$ obtained with the present model. We notice that the star mass increases as the index $n$ grows up. This behaviour can be seen in Figure \ref{fig:400}, where we depict the mass of high density white dwarfs as a function of the index $n$, as obtained from the present simplified model. For pure carbon composition, we found that the obtained model mass varies from $M =0.442 \, M_\odot$ for $n=1$, to $M = 1.134 \, M_\odot$ for $n=400$. Notice that for a carbon white dwarf, the Chandrasekhar limit obtained with a full relativistic treatment of the electron gas is $1.45$ $M_\odot$.     

\begin{table}[htbp]
\begin{center}
\caption{\label{table3} Masses of high density white dwarfs.}
\vspace*{.3cm}
\begin{tabular}{c|c}
\hline
 Index $n$ &  Mass ($M_\odot $)  \\
\hline
 0 & 0.442   \\
 1 &  0.700  \\ 
 2 &  0.819  \\
 3 &  0.889  \\ 
 4 &  0.933  \\
 5 &  0.964  \\
 $ \vdots $  & $ \vdots $ \\
\hline
\end{tabular}
\end{center}
\end{table}

\begin{figure}[htpb]
\centering 
\includegraphics[width=.25\textheight]{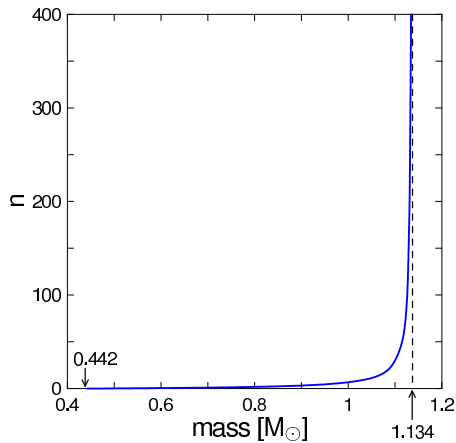}
\vspace*{-.1cm} 
\caption{(Color online) Maximum mass of high density white dwarfs as a function of the model parameter $n$. \label{fig:400}}
\end{figure}

\section{Neutron stars}

Neutron stars are remnant compact objects left by the gravitational collapse of massive stars during a supernova explosion. Neutron stars have typically radius and mass around ten km and 1.5 solar mass, respectively, and are almost entirely composed of   neutrons. It is believed that the mass density at the center of neutron stars may reach or exceed the density of the normal nuclear matter, which is $\rho_0 \approx\ 2.5  \times 10^{14}$ ${\rm g\cdot cm}^{-3}$ \cite{Salaris,BH,Glend,Fridolin}.  

Let us then consider the application of the present simplified model to describe neutron stars entirely composed of free neutrons. In this case, for low densities the neutrons inside the neutron star can be treated in first approximation as a free Fermi gas. Thus, if the neutrons are treated as non-relativistic particles, the polytropic form $P = K_n  \rho^{5/3} $ is still valid, with the proportionality constant now given by $K_n = 5.3802 \times 10^9$ in cgs units \cite{BH}. Therefore, the equations (\ref{radius1}) to (\ref{mass41}) derived in the last section are still formally applicable. For example, for the central density $\rho_c = \ 3.0 \times 10^{14} $ ${\rm g\cdot cm}^{-3}$ and $ n = 2 $, we obtain a neutronic sphere of radius $R = 15.2$ km and mass of $ M = 0.511 \, M_\odot$, which are compatible with typical values of observed neutron stars.

Neutron particles are much heavier than electrons, and become relativistic only for very high densities ($\rho \gg 6.0 \times 10^{15} $ ${\rm g\cdot cm}^{-3}$). So, in the present model the neutron star mass has no upper limit, since the mass is an increasing function of the central density, as given by equation (\ref{mass31}). This is a direct consequence of the use of the Newtonian Gravitation Law in the present simplified model. 

In fact, the high densities present inside neutron stars require that the calculations of the structure of neutron stars as well as their global properties have to be treated in the framework of Einstein's General Theory of Relativity. The resulting structure equations are named the Tolman-Oppenheimer-Volkoff (or simply TOV) equations \cite{BH,TOV1,TOV2}. TOV  equations describe the structure of spherically symmetric bodies in static gravitational equilibrium, which are derived from the solution of the Einstein equations for a time-invariant and spherically symmetric metric, named the Schwarzschild metric. In addition, the equation of state $f(\rho,P)=0$ relating the pressure and the energy density of mater has to be provided, in order to solve the TOV equations. 

\begin{figure}[htpb]
\centering 
\includegraphics[width=.25\textheight]{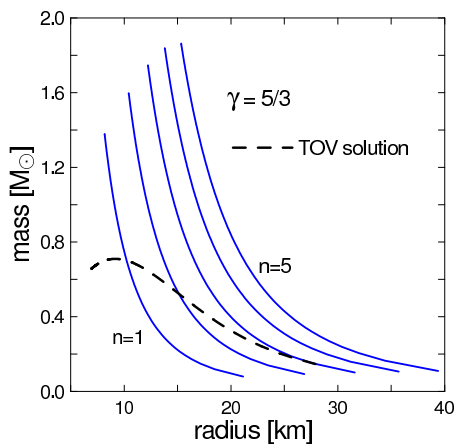}
\vspace*{-.1cm} 
\caption{(Color online) The solid lines represent the neutron star mass-radius relations obtained from the present model. The dashed line stand for the results obtained from the original TOV model. \label{fig:500}}
\end{figure}

\begin{figure}[htpb]
\centering 
\includegraphics[width=.25\textheight]{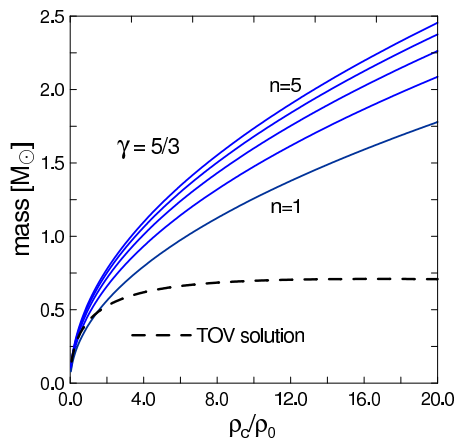}
\vspace*{-.1cm} 
\caption{(Color online) The solid lines represent the neutron star radius as a function of the central density obtained from the present model. The dashed line stand for the original TOV model. \label{fig:600}}
\end{figure}

The structure of low density neutron stars can be derived approximately by the Newtonian Gravitation Law, with the equation of state of the non-relativistic free neutron gas described as a polytropic with the adiabatic index $\gamma = 5/3$.  So, in this framework, the neutron star structure is described by the equilibrium equation (\ref{eql}). When the polytropic equation of state $p(\rho) = K \rho^\gamma$ is assumed for the free neutron gas, then equation (\ref{eql}) leads to the Lane-Emden equation \cite{Salaris,BH}. In this approach, the neutron star radius as a function of the central density can be put in the form \cite{BH} 
\begin{equation}
R=14.64 \left( \frac{\rho _{c}}{10^{15} \,
{\rm g \cdot cm}^{-3}}\right) ^{-1/6}\, {\rm km},  \label{3.48}
\end{equation}%
while the neutron star mass reads
\begin{eqnarray}
M &=&1.102 \left( \frac{\rho_{c}}{10^{15} \,
{\rm g \cdot cm}^{-3}}\right)^{1/2} \, M_{\odot}, \nonumber \\
 &=& \left( \frac{15.12 \, {\rm km}}{R} \right)^{3} \, M_{\odot}.  \label{1.49}
\end{eqnarray}
Notice that equations (\ref{radius1}) to (\ref{mass41}) derived in the last section are the equivalent expressions of the present simplified model for the two previous equations. 

The existence of a maximum value for the mass of a neutron star is manifestly a pure general-relativity effect, which does not occur in  the Newtonian Gravitation. However, the precise value of the mass limit of neutron stars is not yet established, since the nuclear matter equation of state is not well known. Nevertheless, when the mass of the neutron star is greater than the mass limit, the neutron star collapses and becomes a black hole.

If one considers neutron stars almost entirely composed of neutrons, these neutrons can hardly be treated as an ultra relativistic gas, considering the existing densities inside neutron stars. So, a neutron star described by the Newtonian Gravitation Law will always produce stable configurations, since the hydrostatic pressure goes with $\rho^{\gamma}$,  and $\gamma > 4/3$ for free neutrons present inside the neutron star. In other words, this means that the higher the central density, the greater the mass and the smaller the radius of the neutron star. This explains the non existence of a maximum mass for Newtonian neutron stars.

We can see in Figure \ref{fig:500} the mass-radius diagrams obtained from the present simplified model, equation ({\ref{mass41}), for some values of the parameter $n$. The radii as a function of the central density, obtained from equation (\ref{radius1}), are plotted in Figure \ref{fig:600}, for the same set of the parameter $n$. In the figures, are also plotted the solutions of TOV equations (dashed lines) applied to the structure of neutron stars entirely composed of noninteracting neutrons. 

The maximum mass derived from the original TOV equations is only $\approx 0.7\, M_\odot$, when the neutron star is supposed to be entirely composed of noninteracting neutrons at zero temperature \cite{BH,TOV1,TOV2}. This value is known as the Oppenheimer-Volkoff limit, the analogous of the Chandrasekhar limit for white dwarfs. However, more realistic equations of state developed in recent decades provide maximum masses up to $ 2.0 \, M_\odot$. 

The application of the model to the study of neutron stars structure does not accurately reproduces the results provided by the TOV equations. Even though, the obtained masses and radii are comparable with the observed values given in the literature. Both descriptions disagree mainly for high central density neutron stars, but for low  central densities, and especially for $n = 3$ , the models produce equivalent results, as we can see from Figures \ref{fig:500} and \ref{fig:600}.

\section{Conclusions}

In this work we present a simplified model for the calculation of the structure of white dwarfs and neutron stars in hydrostatic equilibrium. In the model, the mass density inside the star is described as a decreasing function of the distance from the center of the  star and parametrized by a positive exponent $n$.

Assuming the validity of Newtonian gravitation, the model provides the exact values of mass and radius of the star as a function of central density, given the exponent $n$. Interestingly, this can be done avoiding any kind of numerical integration, since the   calculations of the compact star structure are provided by simple algebraic relationships. In fact, the calculations involve solely   integration of polynomial functions, which can be easily carried out.

Nevertheless, the results provided by accurate models describing the structure of white dwarfs in hydrostatic equilibrium are roughly reproduced by the present toy model, for appropriate choices of the central density and the parameter $n$.  In particular, for relativistic electrons (adiabatic index $\gamma = 4/3$), the model predicts the existence a maximum possible mass for the white dwarf, which is in good agreement with the Chandrasekhar limit. 
 
 However, the model can hardly reproduce the properties of neutron stars since the correct description of the equilibrium structure of these kind of compact objects imposes the consideration of general-relativity effects, which are not included in the present toy model. We point out that the correct description of the equilibrium properties of neutron stars is provided by the Tolman-Oppenheimer-Volkoff (TOV) equations, which are derived from the Einstein's field equations applied to a time-invariant and spherically symmetric mass distribution. As a purely general-relativity effect, neutron stars must have a maximum possible mass, for finite values of the central density and radius.  For cold ($T=0$) neutron stars entirely composed of free neutrons, the maximum mass is around $0.7 \, M_\odot$, with the radius $\approx 9.6 \, {\rm km}$ and the central density $\rho_c \approx 5.0 \times 10^{15}$ ${\rm g\cdot cm}^{-3}$ \cite{BH,TOV2}.
 
Thus, despite its simplicity, the present model can be used as a simple but helpful approach for a first contact with the stellar astrophysics and the physics of compact stars.

\section{Acknowledgment}
I am grateful to the referee for the valuable suggestions and criticisms added to the final version of the manuscript. I thank the Brazilian foundation CNPq for the financial support.

\end{document}